%% file: main.tex
\documentclass[10pt,journal,compsoc]{IEEEtran}
\usepackage{algorithm}
\usepackage{algpseudocode}
\usepackage[colorinlistoftodos,prependcaption,textsize=tiny]{todonotes}
\usepackage{listings}
\usepackage{booktabs}

\usepackage[hyphens]{url}
\usepackage{hyperref}
\hypersetup{breaklinks=true}

\begin{document}


\title{Kokkos Kernels: Performance Portable Sparse/Dense Linear Algebra and Graph Kernels}


\author{
\IEEEauthorblockN{Sivasankaran Rajamanickam\IEEEauthorrefmark{1},
Seher Acer\IEEEauthorrefmark{1},
Luc Berger-Vergiat\IEEEauthorrefmark{1},
Vinh Dang\IEEEauthorrefmark{1},
Nathan Ellingwood\IEEEauthorrefmark{1},
Evan Harvey\IEEEauthorrefmark{1},
Brian Kelley\IEEEauthorrefmark{1},
Christian R. Trott\IEEEauthorrefmark{1}, 
Jeremiah Wilke\IEEEauthorrefmark{1},
Ichitaro Yamazaki\IEEEauthorrefmark{1}}
\\\IEEEauthorblockA{\IEEEauthorrefmark{1}Sandia National Laboratories\\
\emph{srajama, sacer, lberge, vqdang, ndellin, eharvey, bmkelle, crtrott, jjwilke, iyamaza}@sandia.gov}
}


\algnewcommand{\LineComment}[1]{\State \(\triangleright\) #1}

\newcommand{\shortauthors}{Trott et al.}





\definecolor{commentsColor}{rgb}{0.0, 0.4, 0.2}
\definecolor{keywordsColor}{rgb}{0.6, 0.0, 0.4}
\definecolor{stringColor}{rgb}{0.0, 0.0, 0.7}

\lstset{ %
  backgroundcolor=\color{white},   
  basicstyle=\scriptsize,          
  breakatwhitespace=false,         
  breaklines=true,                 
  captionpos=b,                    
  commentstyle=\color{commentsColor}\textit,    
  keepspaces=true,                 
  keywordstyle=\color{keywordsColor}\bfseries,       
  language=C++,                 
  rulecolor=\color{black},         
  showspaces=false,                
  showstringspaces=false,          
  showtabs=false,                  
  stepnumber=1,                    
  stringstyle=\color{stringColor}, 
  tabsize=2,	                  
  title=\lstname,                  
  columns=fixed                    
}

\IEEEtitleabstractindextext{%
\begin{abstract}
 As hardware architectures are evolving in the push towards exascale, developing Computational Science and Engineering (CSE) applications depend on performance portable approaches for sustainable software development. This paper describes one aspect of performance portability with respect to developing a portable library of kernels that serve the needs of several CSE applications and software frameworks. We describe Kokkos Kernels, a library of kernels for sparse linear algebra, dense linear algebra and graph kernels. We describe the design principles of such a library and demonstrate portable performance of the library using some selected kernels. Specifically, we demonstrate the performance of four sparse kernels, three dense batched kernels, two graph kernels and one team level algorithm. 
\end{abstract}

}


\maketitle

\section{Introduction}





The number of architectures that are of interest to Computational Science and Engineering (CSE) applications has increased rapidly in the past few years. This includes the emergence of ARM based supercomputers such as Astra and Fugaku, and supercomputers with accelerators from AMD, Intel and NVIDIA in production at the same time. 
Kokkos is an ecosystem that serves the needs of CSE applications in this diverse landscape. Kokkos Core \cite{edwards2014kokkos} provides a library-based programming model for performance portability of CSE applications. Several CSE applications rely on  general-purpose libraries for their performance-critical, kernels. Application usage varies from dense matrix operations such as Basic Linear Algebra Subroutines (BLAS) \cite{lawson1979basic} to sparse matrix operations \cite{wang2014intel,naumov2010cusparse}. However, no portable solution existed for sparse/dense linear algebra kernels before the Kokkos Kernels library was created as part of the Advanced Technology Development and Mitigation program and the Exascale Computing Program. Kokkos Kernels was originally developed to provide linear algebra and graph kernels for the Trilinos framework. It later became part of the Kokkos ecosystem to form one portable ecosystem.

The need for portability is different in terms of programming models and libraries for common operations. The complexity in portable programming models is primarily in providing abstractions to parallelize kernels within applications. When designing a portable library with several kernels, the key challenges are in providing: efficient and general interfaces, abstractions at the right levels for several application needs and  portable algorithms for several of these kernels. We will demonstrate use cases for each one of these aspects in the rest of the paper.

Kokkos Kernels has already been adopted as the primary linear algebra layer in software frameworks such as Trilinos and several applications such as SPARC and EMPIRE. This paper introduces the software, clarifies the design choices behind the software, describes few new algorithms and demonstrates their performance on CPUs and GPUs. 

The novel algorithm, data structure and interface contributions in Kokkos Kernels are as follows.
\begin{itemize}
    \item Sparse Linear Algebra: The key contributions of Kokkos Kernels with respect to sparse linear algebra are in developing data structures and portable algorithms, well defined interfaces that allow for good performance on new architectures, interfaces to optimized kernels from vendors when available, and specialized kernels (fused, structure exploiting) where appropriate.
    \item Dense Linear Algebra: Kokkos Kernels provides a hierarchical dense linear algebra implementation that can be used within multiple levels of the parallel hierarchy which is critical for performance in accelerator architectures. Dense linear algebra interfaces support calls that can utilize an entire accelerator, a team of threads, or a single thread within a sequential execution unit. 
    \item Graph Algorithms: Kokkos Kernels provides new algorithms for graph operations that are key for linear algebra operations and for social network analysis.
    \item Batched Data Structure and Utilities: Kokkos Kernels provides data structures and utilities that are parallel context-aware (team, thread-aware) so hierarchical parallel algorithms can be implemented efficiently. 
\end{itemize}

In terms of software design, Kokkos Kernels balances between the following design goals.
\begin{itemize}
    \item Provide a portable, production-ready and high performance library that a higher level framework such as Trilinos \cite{trilinos-website}, PETSc \cite{balay2019petsc}, hypre \cite{falgout2002hypre}, and applications such as ExaWind and SPARC can build on.
    \item Serve as a research vehicle where new portable algorithms can be easily developed and deployed.
    \item Provide a common interface to high-performance, architecture-specific kernels from libraries such as cuBLAS \cite{naumov2010cusparse}, rocSPARSE \cite{rocsparse} and Intel MKL \cite{Intel_MKL} when they exist alongside our native implementations. 
\end{itemize}

\noindent

The rest of the paper is organized as follows. Sparse and dense linear algebra features are described in Sections \ref{sec:sparse} and \ref{sec:dense} respectively. Portable graph algorithm features are described in Section \ref{sec:graph}. We do not demonstrate performance of all the features of the Kokkos Kernels library in this paper. This has been shown in other papers and we cite those papers where appropriate. New and previously unpublished results are included here. 
Unless noted otherwise, the performance results shown in this paper\footnote{Instructions for reproducing performance results are available at \url{https://github.com/kokkos/kokkos-kernels/wiki}.} were collected using the 3.3.01 versions of Kokkos and Kokkos Kernels compiled with double precision scalars, GCC C++ version 7.2.0 and cuda 10.1.105.

\textbf{Summary of New Results}:  We summarize the new algorithms and results that are included in this paper here. We show results for four sparse matrix algorithms, a structured SpMV (\ref{sec:structured-spmv}), a fused Jacobi-SpGEMM (\ref{sec:jacobispgemm}), sparse matrix-matrix addition (\ref{sec:spadd}), and a supernodal sparse triangular solve (\ref{sec:sn_sptrsv}). We demonstrate the portable performance of our team-level batched routines on GPUs (\ref{sec:batched-blas}), which were previously demonstrated only on CPUs. We cover graph kernels for distance-2 coloring (\ref{sec:d2coloring}) and distance-2 maximum independent set (\ref{sec:d2mis}), and show performance on GPUs. Finally, we demonstrate the performance of bitonic sorting in comparison to Thrust \cite{Thrust}, CUB \cite{merrill2015cub}, and Kokkos \cite{edwards2014kokkos}.

\section{Sparse Linear Algebra}
\label{sec:sparse}

For its sparse linear algebra implementations, Kokkos Kernels uses the compressed row sparse (CRS) format to store sparse matrices, which is commonly referred to as \emph{CrsMatrix}. That format is widely used in sparse linear algebra libraries (Trilinos, PETSc, hypre) and is the most commonly used sparse matrix format in CSE software and applications.  Most importantly this allows Kokkos Kernels and Trilinos to interface easily as Tpetra (Trilinos distributed sparse linear algebra package) also relies on that matrix format. The concept of vectors is not formally introduced in Kokkos Kernels, it simply relies on Kokkos views of rank 1 and rank 2 to represent a single vector or multiple vectors.
The row pointer and column indices of the CrsMatrix form the graph of the matrix and are implemented in the \texttt{StaticCrsGraph} associated with the matrix. The \texttt{StaticCrsGraph}  defines the row pointer as constant after the matrix has been created; this avoids the potential for memory reallocation by the user, which is typically costly on GPU architectures.

In the rest of this section the following core kernels for sparse linear algebra are presented: sparse matrix vector multiplication (SpMV), sparse matrix matrix addition (SpAdd), sparse matrix matrix multiplication (SpGEMM) and sparse triangular solver (SpTRSV). The last three kernels use an interface that splits the symbolic operations from numeric operations for better performance.
\subsection{SpMV: Sparse Matrix Vector Multiply}
\input{sparse-spmv}

\subsection{SpGEMM: Sparse Matrix Matrix Multiply}
\input{sparse-spgemm}

\subsection{SpAdd: Sparse Matrix Addition}
\input{sparse-spadd}

\subsection{Sparse Triangular Solver}
\input{sparse-sptrsv}

\section{Dense Linear Algebra}
\label{sec:dense}

\subsection{BLAS/LAPACK Interfaces}
When using BLAS/LAPACK functions, users tend to rely on vendor library functions which are well optimized for vendor-specific hardware architectures. Commonly, users need to write different function interfaces for different computing platforms, such as Intel's MKL vs. NVIDIA's cuBLAS. Each BLAS/LAPACK vendor library only supports its built-in real/complex data types and column/row major data layouts. Although these functions are highly optimized, in practice, high performance is usually obtained from large problem sizes. The motivations of Kokkos Kernels BLAS/LAPACK interfaces are (i) to provide a single interface to vendor BLAS libraries on heterogenous computing platforms; (ii) to support custom user-defined scalar data types e.g., Automatic Differentiation, Ensemble data types for uncertainty quantification \cite{phipps2017embedded},  with Kokkos native implementation; (iii) to provide customized performance solution for certain problem sizes; (iv) to explore new performance oriented interfaces

Kokkos Kernels provides interfaces for most BLAS-1 type functions, BLAS-2 type function (\textsc{gemv} - matrix vector multiplication), a subset of BLAS-3 type functions (\textsc{gemm} - matrix matrix multiplication, \textsc{trsm} - triangular linear system solve with multiple right-hand-sides, and \textsc{trmm} - triangular matrix multiplication), and a subset of LAPACK functions (\textsc{gesv} - linear equation system solve using LU factorization, \textsc{trtri} - triangular matrix inverse). Kokkos Kernels' APIs are implemented in a simple, generic way so that the called functions are able to run on a wide range of architectures. These functions accept \textit{Kokkos::View}s instead of raw pointers as inputs and outputs. Depending on where data resides, Kokkos Kernels calls the right functions for the targeted backend. Additionally, Kokkos Kernels' BLAS-1 type functions are also able to operate on multi-vectors, where the 1D BLAS operations are applied to each vector respectively. 

The default performance-portable implementations for BLAS/LAPACK functions in Kokkos Kernels employ Kokkos execution patterns and policies. Specialized TPL implementations pointing to vendor-optimized BLAS/LAPACK are also provided when appropriate. Using the approach described in section \ref{sec:software_design}, the BLAS/LAPACK interfaces enable straightforward, convenient calls to vendor libraries. Kokkos Kernels supports host BLAS/LAPACK TPLs such as Intel's MKL, IBM's ESSL, OpenBLAS, etc.. For GPU architectures, the currently supported TPLs are cuBLAS and MAGMA (LAPACK functionalities). There are custom BLAS/LAPACK kernels implemented for performance reason as well. For instance, the dot-based \textsc{gemm} implements the optimization for $C = \beta*C + \alpha*A^T*B$ where A, B matrices are both tall and skinny, and C matrix is small. In this particular implementation, each entry of C is computed by performing the dot product of respective columns of A and B matrices. It is noted that since the dot products are performed on very long vectors, each dot product is assigned to a team and threads within a team are collectively perform the assigned dot product. The advantage of such an implementation for orthogonalization kernels in eigen solvers has been shown in the past \cite{sphynx}.

\subsection{Parallel Batched BLAS/LAPACK Interfaces}
\label{sec:batched-blas}
The Kokkos Kernels batched BLAS/LAPACK interface provides multiple functor-level interfaces for dense linear algebra (DLA), which is suitable for Kokkos hierarchical parallelism. Unlike other batched BLAS and LAPACK interface, Intel batched \textsc{gemm} \cite{Intel_MKL}, cuBLAS batched \textsc{gemm} \cite{Nvidia_cuBLAS}, MAGMA batched \textsc{gemm} \cite{MAGMA_Batched}, we do not provide a front-level (or subroutine) interface that launches a streaming parallel kernel. Instead, we provide a functor-level interface that can be used in a Kokkos parallel execution patterns e.g., parallel\_for, parallel\_reduce and parallel\_scan. The advantage of this approach is that a user can compose various batched DLA subroutines and exploit temporal locality via the functor-level interfaces. Vectorized versions for CPU and Intel Knights Landing architectures have been studied in the past \cite{kim2017designing}.

Most vendor-provided DLA libraries such as Intel MKL and NVIDIA cuBLAS perform well for large problem sizes. For small problem sizes, it is not feasible to use vendor-optimized DLA libraries as even a function call or error check already puts quite an amount of overhead for such problems. Furthermore, cuBLAS cannot be nested inside of a parallel region. The main difference from other vendor-provided DLA libraries is that Kokkos Kernels batched APIs are very \textit{light-weight} generic implementations focusing on small matrix sizes; kernels are developed and tuned from the application context. Kokkos batched APIs provide generic functor-level implementations that can be embedded in a parallel region. Hence, a single parallel\_for can be launched to compute a sequence of DLA operations such as \textsc{gemm}, \textsc{lu} and \textsc{trsv}. Moreover, the batched APIs can be mapped to Kokkos hierarchical parallelism and also provide various implementations of algorithms with a template argument, which allows users to choose the batched routines best suited for their application context.

Kokkos Kernels provides three choices of functor-level interfaces for batched BLAS/LAPACK DLA subroutines: \textit{Serial}, \textit{Team}, and \textit{TeamVector}. The \textit{Serial} (Algorithm \ref{alg:BatchedSerial}) interface maps to a single Kokkos thread and uses serial execution patterns internally. The \textit{Team} (Algorithm \ref{alg:BatchedTeam}) interface maps to a team of Kokkos threads and uses the Kokkos \textit{parallel\_for} execution pattern to assign a Kokkos thread to a subset of the subroutine's work internally, for example, one Kokkos thread per dot product. The \textit{TeamVector} (Algorithm \ref{alg:BatchedTeamVector}) interface maps to a team of Kokkos threads and one or more vector lanes per Kokkos thread; \textit{TeamVector} uses two Kokkos \textit{parallel\_for} execution patterns internally: one to assign a Kokkos thread to a subset of the subroutine's work and a second to assign a vector lane to a subset of the subroutine's work. Lastly, Kokkos Kernels provides a SIMD data type intended to be used for allocating Kokkos views that are passed to the \textit{Serial} interface or \textit{Team} interface (Algorithm \ref{alg:BatchedTeamWithSimd}); this SIMD type affects the data access pattern of the subroutine and provides a mechanism for reducing GPU memory transactions via double word loads and stores. Depending on the application's matrix sizes and required batched DLA subroutines, the user will choose what is suitable for their needs.
\begin{figure}
    \centering
    \includegraphics[scale=0.8]{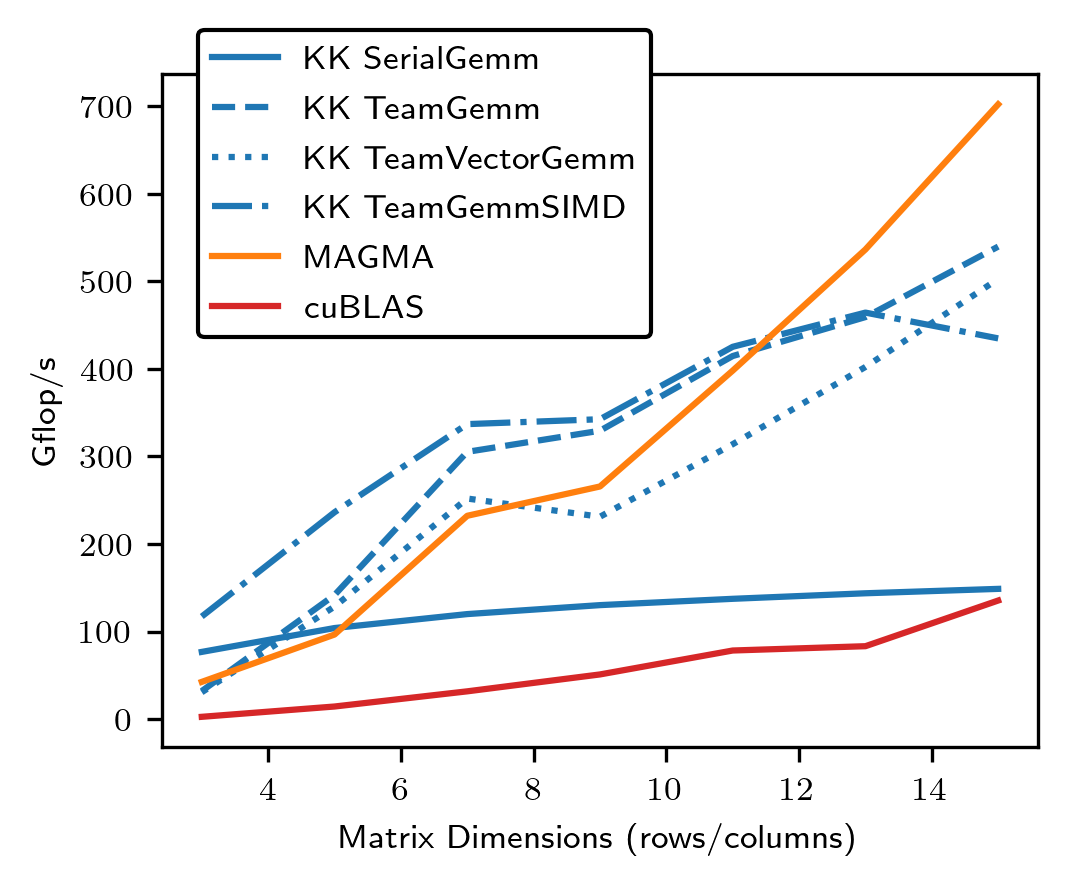}
     \vspace{-0.4cm}
    \caption{Batched gemm on small matrices: Kokkos Kernels vs. MAGMA vs. cuBLAS}
    \label{fig:gemm_kokkos_magma_cublas}
     \vspace{-0.5cm}
\end{figure}

Figure \ref{fig:gemm_kokkos_magma_cublas} shows the performance of Kokkos Kernels (using all four interfaces mentioned above) against MAGMA, and cuBLAS for batched \textsc{gemm} on small
matrices. We evaluate the performance for random-valued, square matrices
of sizes 3, 5, 7, 9, 11, 13, and 15, with a batch size of 163,840. All Kokkos Kernels (KK) batched \textsc{gemm} experiments average times over 2000 trials. The KK, MAGMA, and cuBLAS comparisons use O3 optimization on an NVIDIA Tesla V100 GPU. This data was collected with MAGMA 2.5.4 batched \textsc{gemm} and the Kokkos Kernels develop branch.
It is observed that all Kokkos Kernels implementations outperform the cuBLAS batched \textsc{gemm}. For larger sizes of 13 and 15, MAGMA performs better as compared to the KK implementations. For example, MAGMA achieves 1.1x, 1.6x speedups w.r.t. KK SIMD Team Gemm interface for these block sizes, respectively. For smaller sizes of 3, 5, 7, 9, 11, KK SIMD Team Gemm interface is faster than MAGMA with the speedups of 2.8x, 2.4x, 1.5x, 1.3x, 1.1x, respectively.


\begin{figure}
    \centering
    \includegraphics[scale=0.8]{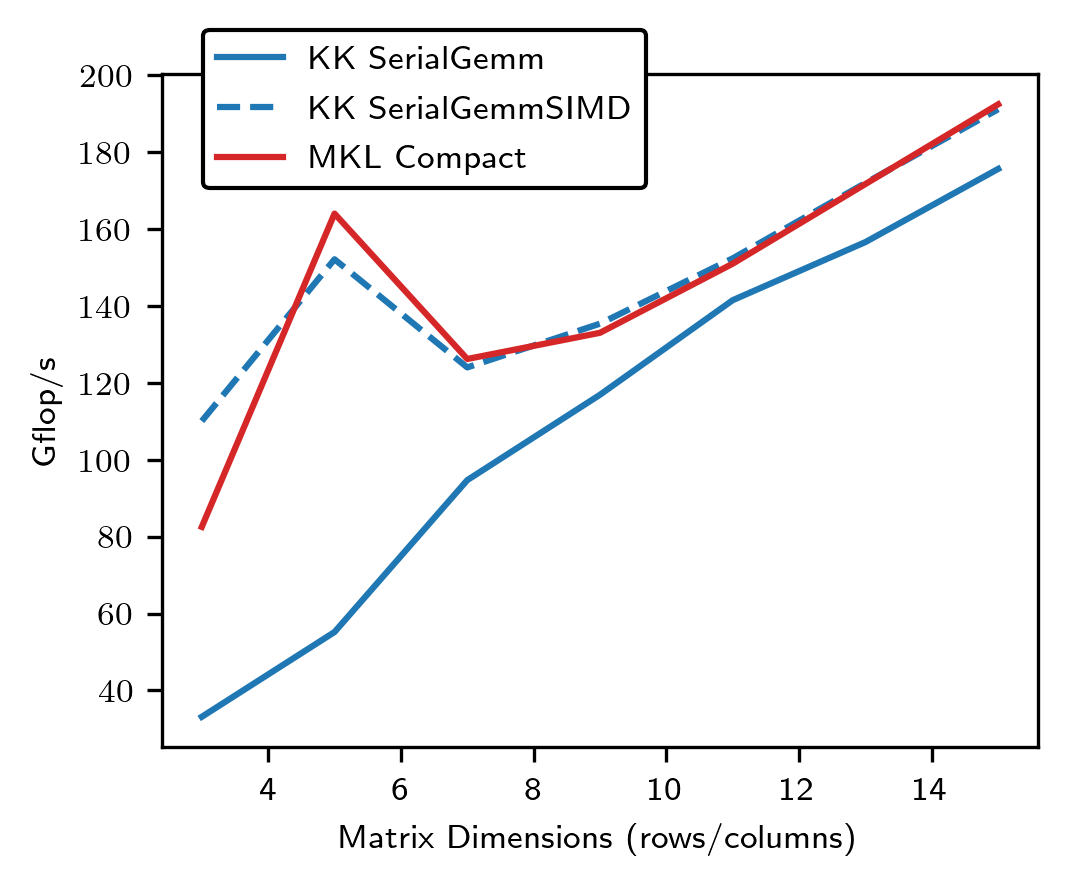}
     \vspace{-0.4cm}
    \caption{Batched gemm on small matrices: Kokkos Kernels vs. MKL}
    \label{fig:gemm_kokkos_mkl}
     \vspace{-0.5cm}
\end{figure}
To evaluate the batched \textsc{gemm} performance on CPU backend, we carry out the same experiment as with the aforementioned GPU case and compare against the MKL compact implementation, as shown in Figure \ref{fig:gemm_kokkos_mkl}. The KK and MKL comparisons use O3 optimization with Intel C++ version 19.0.3.199, with 96 OpenMP threads on a node with two Intel Xeon Platinum 8160 (Skylake) processors. We only show the KK Serial interfaces' results since these are the best performance number over the other KK implementations. Using SIMD type, we can achieve an average speedup of 1.7x as compared to KK Serial interface for all matrix sizes. We see that the performance of MKL Compact is comparable to KK Serial interface with SIMD type.

\begin{figure}
    \centering
    \includegraphics[scale=0.8]{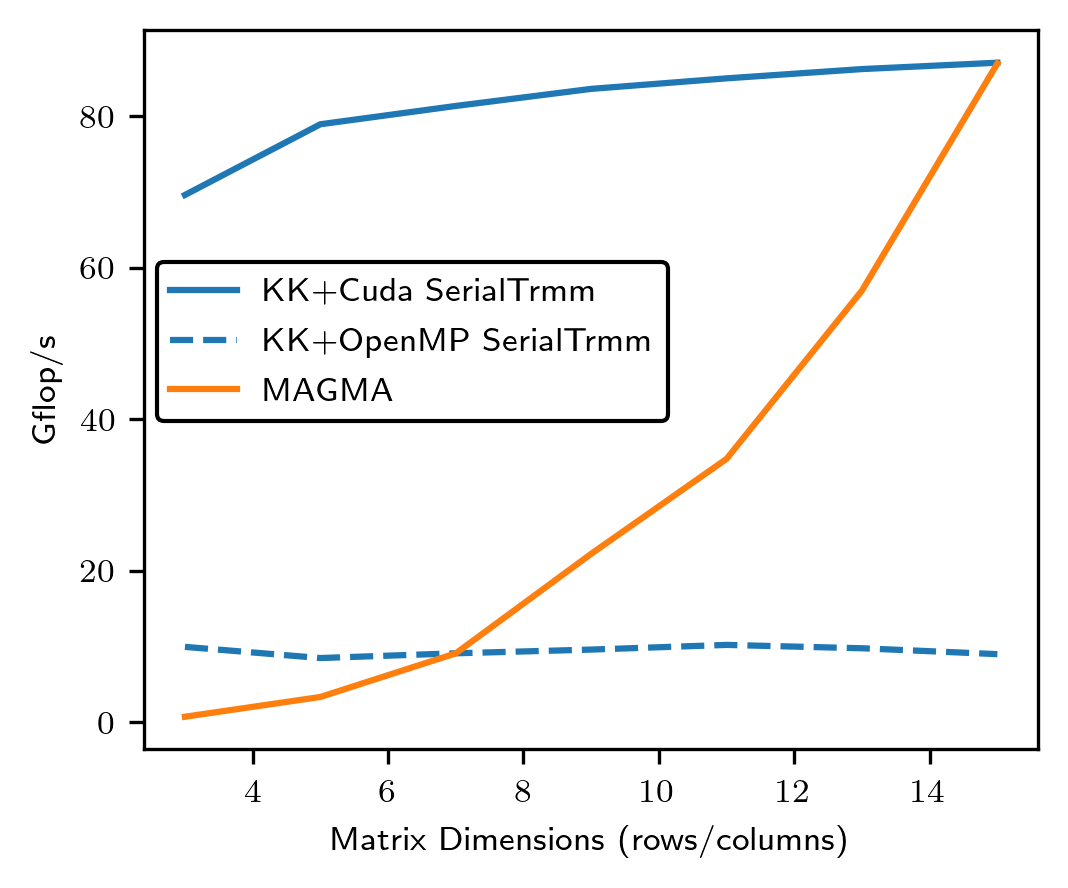}
    \caption{Batched trmm on small matrices: Kokkos Kernels vs. MAGMA}
     \vspace{-0.4cm}
    \label{fig:trmm_kokkos_magma}
     \vspace{-0.3cm}
\end{figure}

Figure \ref{fig:trmm_kokkos_magma} compares Kokkos Kernels with MAGMA for batched \textsc{trmm} on small matrices. The input matrices are square with a batch size of $2^{19}$ and are populated with random values.
All Kokkos Kernels batched \textsc{trmm} experiments average times over 2000 trials. The KK+Cuda and MAGMA comparisons use O3
optimization on an NVIDIA Tesla V100 GPU. The KK+OpenMP
comparison uses O3 optimization with 96 OpenMP threads on a node with two Intel Xeon Platinum 8160 (Skylake) processors. This data was collected with MAGMA 2.5.4 batched \textsc{trmm} and the Kokkos Kernels develop branch. cuBLAS and MKL batched \textsc{trmm} performance is omitted since these BLAS/LAPACK libraries do not provide a batched \textsc{trmm} interface. Figure \ref{fig:trmm_kokkos_magma} shows that KK+Cuda outperforms both KK+OpenMP and MAGMA. For matrix sizes of 3, 5, 7, 9, 11, 13, the KK+Cuda SerialTrmm interface is faster than MAGMA with speedups of 95.4x, 23.6x, 8.9x, 3.8x, 2.4x, 1.5x, respectively. For the matrix dimension of 15, MAGMA is comparable to KK+Cuda.

\begin{figure}
    \centering
    \includegraphics[scale=0.8]{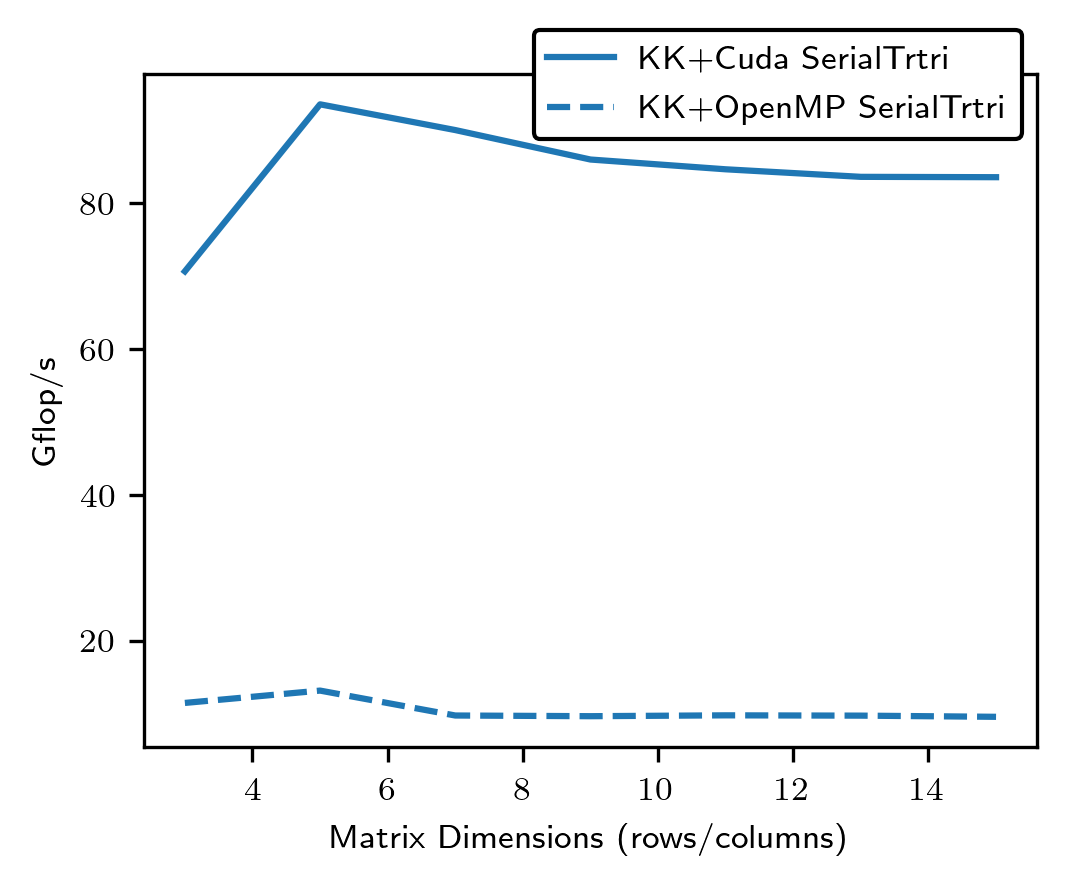}
     \vspace{-0.4cm}
    \caption{Batched trtri on small matrices: Kokkos Kernels}
    \label{fig:trtri_kokkos}
     \vspace{-0.2cm}
\end{figure}

Figure \ref{fig:trtri_kokkos} compares Kokkos Kernels with the Kokkos Cuda backend and Kokkos OpenMP backend for batched \textsc{trtri} on small matrices. The input matrices are square with a batch size of $2^{19}$ and are populated with random values.
All Kokkos Kernels batched \textsc{trtri} experiments average times over 2000 trials. The KK+Cuda comparisons use O3 optimization on an NVIDIA Tesla V100 GPU. The KK+OpenMP comparison uses O3 optimization with 96 OpenMP threads on a node with two Intel Xeon Platinum 8160 
(Skylake) processors. This data was collected with the Kokkos Kernels develop branch. MAGMA, cuBLAS and MKL batched \textsc{trtri} performance is omitted since these BLAS/LAPACK libraries do not provide a batched \textsc{trtri} interface. Figure \ref{fig:trtri_kokkos} shows that KK+Cuda outperforms KK+OpenMP.

\begin{algorithm}[t]
\caption{Batched Serial}
\label{alg:BatchedSerial}
\begin{algorithmic}[1]
\small
    \Procedure{BatchedSerial}{Amats, Bmats}
        \LineComment{Using Kokkos RangePolicy}
        \ForAll{$A, B \in Amats, Bmats$}
            \State $A \gets SerialTrtri(A)$
            \State $A \gets SerialTrmm(\alpha, A, B)$
        \EndFor
    \State\Return{$A$}
    \EndProcedure
\end{algorithmic}
\end{algorithm}


\begin{algorithm}[t]
\caption{Batched Team}
\label{alg:BatchedTeam}
\begin{algorithmic}[1]
\small
    \Procedure{BatchedTeam}{Amats, Bmats, Cmats}
        \LineComment{Using Kokkos TeamPolicy}
        \ForAll{$A, B, C \in Amats, Bmats, Cmats$}
            \State $C \gets TeamGemm(\alpha, A, B, \beta, C)$ 
        \EndFor
    \State\Return{$C$}
    \EndProcedure
\end{algorithmic}
\end{algorithm}


\begin{algorithm}[t]
\caption{Batched TeamVector}
\label{alg:BatchedTeamVector}
\begin{algorithmic}[1]
\small
    \Procedure{BatchedTeamVector}{Amats, Bmats, Cmats}
        \LineComment{Using Kokkos TeamPolicy}
        \ForAll{$A, B, C \in Amats, Bmats, Cmats$}
            \State $C \gets TeamVectorGemm(\alpha, A, B, \beta, C)$
        \EndFor
    \State\Return{$C$}
    \EndProcedure
\end{algorithmic}
\end{algorithm}


\begin{algorithm}[t]
\caption{Batched SIMD}
\label{alg:BatchedTeamWithSimd}
\begin{algorithmic}[1]
\small
    \LineComment{Inputs are SIMD Kokkos Views. ThreadVectorRange within TeamPolicy is "unconventional" Kokkos Hierarchical Parallelism}
    \Procedure{BatchedSimd}{Amats, Bmats, Cmats}
        \LineComment{Using Kokkos TeamPolicy}
        \ForAll{$Av, Bv, Cv \in Amats, Bmats, Cmats$}
            \LineComment{Using Kokkos ThreadVectorRange}
            \ForAll{$A, B, C \in Av, Bv, Cv$}
                \State $C \gets SerialGemm/TeamGemm(\alpha, A, B, \beta, C)$
            \EndFor
        \EndFor
    \State\Return{$C$}
    \EndProcedure
\end{algorithmic}
\end{algorithm}
\vspace{-0.5 cm}


\section{Graph Algorithms}
\label{sec:graph}

\subsection{Distance-1 Graph Coloring}
Distance-1 graph coloring is the classical problem for undirected graphs: label each vertex with a color
such that no two adjacent vertices have the same color. Minimizing the number of colors is intractable (NP-complete), but an approximate solution is still useful and can be found efficiently by greedy algorithms. Kokkos Kernels has two parallel greedy algorithms for coloring: vertex-based (VB) and edge-based (EB) \cite{MehmetColoring}. Both algorithms are speculative. They assign colors to vertices without thread synchronization, potentially creating an invalid coloring due to data races. Conflicts are then resolved
by uncoloring one endpoint of each edge between two like colors. This process is repeated until the conflict resolution pass finds no conflicts.

The difference between VB and EB is in the granularity of each thread's task.
In VB, each thread loops over a vertex's neighbors to determine
the smallest valid color, and to detect conflicts. In EB, each thread processes only a single edge at a time. In the coloring pass, the color of each endpoint is passed to the other to find the set of valid colors. In the conflict detection pass, if the colors of each endpoint are the same, the endpoint with the larger index is uncolored. VB is usually the faster algorithm in practice, but for irregular graphs with some vertices of high degree (e.g power-law graphs), EB can be much faster on GPUs.

\subsection{Distance-2 Graph Coloring}
\label{sec:d2coloring}
Distance-2 graph coloring and bipartite graph partial coloring (BGPC) are two other closely related coloring problems. Distance-2 coloring is the same as distance-1 coloring, except that any two vertices connected by a path of two or fewer edges must have different colors. Bipartite graph partial coloring is the same, but defined for bipartite graphs and only one part (left or right) of the graph is assigned colors. For undirected (symmetric) graphs with all self-loops, these problems are equivalent. In Kokkos Kernels, the bipartite graph is still represented as a sparse adjacency matrix, where each row is a vertex in the left part, and each column is a vertex in the right part.

For these problems, Kokkos Kernels implements a parallel algorithm called net-based (NB) coloring by Taş et al. \cite{D2Coloring}. The simplest approach to greedy distance-2 coloring is to extend the
distance-1 VB algorithm to loop over both neighbors and neighbors-of-neighbors. However, this nested loop has a high performance cost. Net-based coloring avoids the nested loop by gathering used colors from immediate neighbors twice. After the $k$-th gather pass, each vertex is aware of all used colors in a radius-$k$ neighborhood. If $\Delta$ is the maximum degree of
the graph, this reduces the complexity of each coloring pass from $O(\Delta^2|V|)$ to $O(\Delta|V|)$. In practice, using pure NB coloring is not always faster than VB. NB must recompute the forbidden colors for all vertices after every iteration, while VB only computes them for the uncolored vertices.

\begin{table}[t]
\caption{VB vs. NB performance on selected graphs (NVIDIA V100)}
\begin{tabular}{lrrrrr}
\toprule
graph     & vertices & $\Delta_{avg}$ & $\Delta_{max}$ & VB time & NB time \\ \hline
ecology2   & 999999   & 4.996           & 5              & 0.0101  & 0.0706  \\
af\_shell7 & 504855   & 34.8            & 40             & 0.271   & 0.358   \\
Fault\_639 & 638802   & 44.8            & 318            & 0.612   & 1.172   \\
circuit\_4 & 80209    & 3.84            & 6750           & 210.9   & 21.3 \\
\bottomrule
\end{tabular}
\label{tab:d2}
\end{table}

Table \ref{tab:d2} compares VB and NB performance on some selected graphs from the SuiteSparse collection \cite{Davis2011}. $\Delta_{avg}$ and $\Delta_{max}$ are the average and maximum degrees, respectively. Generally, as $\Delta_{max}$ increases, NB performs better relative to VB.

\subsection{Distance-2 Maximal Independent Set}\label{sec:d2mis}
Distance-2 maximal independent set (MIS-2) is the problem of finding a subset of vertices such that no two vertices are within 2 edges of each other, and where no additional vertex can be added to the set without violating this property. MIS-2 is useful for graph coarsening, especially in algebraic multigrid \cite{BellMIS2}. Each coarse vertex may be formed from a ``root'' vertex in the MIS-2 and its radius-2 neighborhood.

Kokkos Kernels has an optimized, deterministic MIS-2 implementation based on an algorithm by Bell et al. \cite{BellMIS2}, as well as MIS-2 based coarsening. We compare the performance of Kokkos Kernels MIS-2 against the CUSP library \cite{cusp}, also by Dalton and Bell. 15 graphs from SuiteSparse \cite{Davis2011} are used, as well as a $100^3$ 3D Laplacian problem and a $60^3$ 3D elasticity problem generated by MueLu \cite{muelu}. The number of vertices range from 505K to 1.498M, and the average degrees are between 5 and 82. Kokkos Kernels averages a 6x speedup by using prefix-sum worklists, re-randomizing the priorities each iteration, and by representing vertex states more efficiently.

\section{Multi-Level Bitonic Sorting}
\label{sec:utils}
Kokkos Kernels provides a flexible implementation of the bitonic sorting algorithm. Although bitonic sort has a high asymptotic complexity of $O(n\mathit{log}^2(n))$, it has several advantages. It is comparison-based and in-place, unlike e.g. radix sort. Being comparison based means that any comparison function can be used. Bitonic sorting is also highly parallel and  has a memory access pattern friendly to GPUs: the SIMD lanes of each Kokkos thread (i.e. CUDA warp) always access a contiguous range of elements simultaneously.

Kokkos Kernels provides team-level and device-level bitonic sort. Like batched BLAS, the team-level sort can be called from inside a Kokkos parallel kernel and exploits all available parallelism in the team.

\begin{figure}[t]
    \centering
    \includegraphics[scale=0.85]{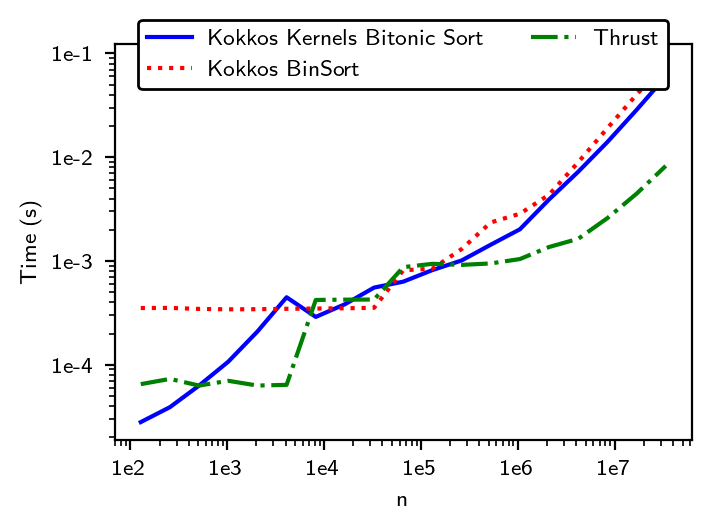}
     \vspace{-0.5cm}
    \caption{Device-level sorting: Kokkos Kernels vs. Thrust and Kokkos}
    \label{fig:devsorting}
     \vspace{-0.4cm}
\end{figure}

Figure \ref{fig:devsorting} compares three device-level sorting implementations on an NVIDIA V100: Kokkos Kernels (bitonic sort), Kokkos (bin sort) \cite{edwards2014kokkos} and Thrust \cite{Thrust}. The input arrays contain $n$ random 32-bit integers, with times averaged over 1000 different arrays. Thrust is obviously the fastest for the device-level sort, but Kokkos Kernels gives an average speedup of 1.3x over Kokkos for $2^{16} \leq n \leq 2^{25}$.

\begin{figure}[t]
    \centering
    \includegraphics[scale=0.85]{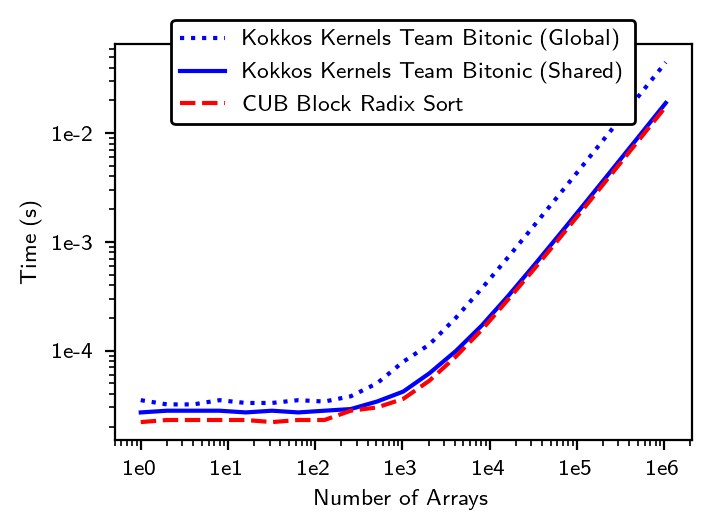}
     \vspace{-0.5cm}
    \caption{Team-level sorting: Kokkos Kernels vs. CUB}
    \label{fig:teamsorting}
    \vspace{-0.4cm}
\end{figure}

Figure \ref{fig:teamsorting} compares Kokkos Kernels against CUB's BlockRadixSort \cite{merrill2015cub} for team-level sorting of $n$ 256-element integer arrays, using teams of 128 threads. An NVIDIA V100, CUDA 11.0 and the built-in CUB were used. Kokkos Kernels can either perform the sorting directly in global memory, or it can process the array in shared memory for a 1.75x speedup. When Kokkos Kernels uses shared memory, it is only 8\% slower than CUB for $n \geq 4096$. All reported times are for the entire kernel, including the transfers between global and shared memory. Although Kokkos Kernels is slightly slower than CUB, it can sort elements of any type, use any comparison function, and is not constrained by hardware registers or shared memory.

\input{softwaredesign.tex}

\section{Acknowledgments}

We would like to acknowledge the contributions of several ``alumni''  and friends of the Kokkos Kernels library development team who have contributed to the library in the past. Especially, Andrew Bradley, Mehmet Deveci, Simon Hammond, Mark Hoemmen, Kyungjoo Kim, and William Mclendon have contributed to the software design and specific algorithmic aspects as can be seen in the algorithm description papers cited here.
Sandia National Laboratories is a multimission laboratory managed and operated by National Technology \& Engineering Solutions of Sandia, LLC, a wholly owned subsidiary of Honeywell International Inc., for the U.S. Department of Energy’s National Nuclear Security Administration under contract DE-NA0003525. SAND2021-3421 O

\bibliographystyle{IEEEtran}
\bibliography{references}

%


\begin{IEEEbiographynophoto}
{Sivasankaran Rajamanickam} is a 
Principal Member of Technical Staff at Sandia National Laboratories. He has a Ph.D. in Computer Science and Engineering from University of Florida (2009). His interests are in high performance computing and sparse linear algebra/solvers. He leads the Kokkos Kernels library and the Trilinos Solver Product.
\vspace{-10 mm}
\end{IEEEbiographynophoto}

\begin{IEEEbiographynophoto}{Seher Acer}
is a Postdoctoral Appointee at Sandia National Laboratories. She has a B.S., M.S., and Ph.D. in Computer Science from Bilkent University. She is interested in high performance computing and combinatorial scientific computing. 
\vspace{-10 mm}
\end{IEEEbiographynophoto}

\begin{IEEEbiographynophoto}{Luc Berger-Vergiat} is a
Limited Term Employee at Sandia National Laboratories. He has a Ph.D. in Civil Engineering from Columbia University (2015). He is interested in parallel sparse linear algebra and sparse linear solvers.
\vspace{-10 mm}
\end{IEEEbiographynophoto}

\begin{IEEEbiographynophoto}{Vinh Dang} is a Senior Member of Technical Staff at Sandia National Laboratories. He has a Ph.D. in Electrical Engineering from the Catholic University of America (2015). His interests are in high performance computing and parallel dense linear algebra/solvers.
\vspace{-10 mm}
\end{IEEEbiographynophoto}

\begin{IEEEbiographynophoto}{Nathan Ellingwood}
is a senior member of technical staff at Sandia National Laboratories. Nathan earned a PhD in Applied Mathematics and Computational Sciences from the University of Iowa. At Sandia he contributes to the Kokkos Core and Kokkos Kernels projects, with a particular focus on testing and release infrastructure.
\vspace{-10 mm}
\end{IEEEbiographynophoto}

\begin{IEEEbiographynophoto}{Evan Harvey}
is a Limited Term Employee at Sandia National Laboratories. He has a B.S. in Computer Science from New Mexico Institute of Mining and Technology (2016). He contributes to the Kokkos Core and Kokkos Kernels projects with a focus on parallel dense linear algebra, software engineering, and continuous integration testing.
\vspace{-10 mm}
\end{IEEEbiographynophoto}

\begin{IEEEbiographynophoto}{Brian Kelley}
is a Member of Technical Staff at Sandia National Laboratories. He has a B.S. in Computer Science and Applied Math from Texas A\&M University (2019). His interests are in shared-memory parallel data structures and algorithms.
\vspace{-10 mm}
\end{IEEEbiographynophoto}

\begin{IEEEbiographynophoto}{Christian Trott}
is a principal member of technical staff at Sandia National Laboratories, where he has worked since acquiring a PhD in Theoretical Physics at TU Ilmenau, Germany. He leads the Kokkos Core project and represents Sandia at the ISO C++ committee. 
\vspace{-10 mm}
\end{IEEEbiographynophoto}

\begin{IEEEbiographynophoto}{Jeremiah Wilke}
is a principal member of technical staff at Sandia National Laboratories.  Jeremiah received his PhD in computational chemistry from the University of Georgia. He has contributed to several HPC efforts including architecture simulation, interconnect design, distributed runtime systems, and novel programming models. His work on Kokkos and Kokkos Kernels has primarily focused on scalable, portable build system design and deployment on diverse HPC systems.
\vspace{-10 mm}
\end{IEEEbiographynophoto}

\begin{IEEEbiographynophoto}{Ichitaro Yamazaki} is a
Member of Technical Staff at Sandia National Laboratories. He has a Ph.D. in Computer Science from University of California, Davis (2008) and is  interested in parallel sparse linear algebra and sparse linear solvers.
\end{IEEEbiographynophoto}\vfill

\end{document}

%% file: sparse-spmv.tex
The Sparse Matrix Vector multiplication (SpMV) is probably one of the most common kernels in sparse linear solvers, for instance to compute residuals, to apply polynomial preconditioning or to project vectors onto new spaces as is performed in domain decomposition and multigrid techniques. SpMV implements the following operation: $y=\beta y + \alpha A x$ where $A$ is a $m\times n$ CrsMatrix, $x$, resp. $y$, is $k$ vectors of length $n$, resp. $m$ and $\alpha$ and $\beta$ are scalar parameters. As is common, four modes are supported for this kernel: plain, transpose, conjugate and conjugate transpose, which are applied to the matrix during the computation.

While providing a unique interface for all modes and for single vs multiple vector multiplication is the most convenient for users, the underlying implementation uses four specialized kernels for performance:
\begin{itemize}
    \item plain or conjugate single vector,
    \item plain or conjugate multiple vectors,
    \item transpose or conjugate transpose single vector,
    \item transpose or conjugate transpose multiple vectors.
\end{itemize}
Furthermore, the implementation is specialized for host execution where a single level of parallelism is used and for GPU execution where three levels of parallelism are employed. These ensure good performance of the kernels on multiple architectures. 
Using the mechanism described in section \ref{sec:software_design}, the native implementation provided by the library can be overridden in favor of specialized TPL implementations. Currently, only cuSPARSE and MKL are supported as TPLs for SpMV. Early performance results of the SpMV kernel can be found in \cite{spmv_report} where the discussion focused on the optimization of the kernel for the OpenMP backend on Intel Knights Landing and IBM Power8 platforms.

\subsubsection{SpMV on structured grids}
\label{sec:structured-spmv}
A variant of the SpMV algorithm for structured grid problems is also available for applications that can represent their problems on structured grids and can thus leverage more computationally efficient algorithms, especially on GPUs.

The algorithm requires the user to provide the dimensions of the underlying box as well as the stencil type used for the discretization (currently only 3pts, 5pts, 7pts, 9pts and 27pts stencils are supported). Based on the box dimensions, the algorithm excludes rows corresponding to points on the skin of the box as these might contain boundary conditions that are treated with the standard SpMV algorithm. For the remaining rows, the algorithm exploits the structure of the stencil to avoid indirection when accessing the entries of $x$ and to select optimal kernel launch parameters based on the stencil length.

\begin{figure}
    \centering
    \includegraphics[scale=1]{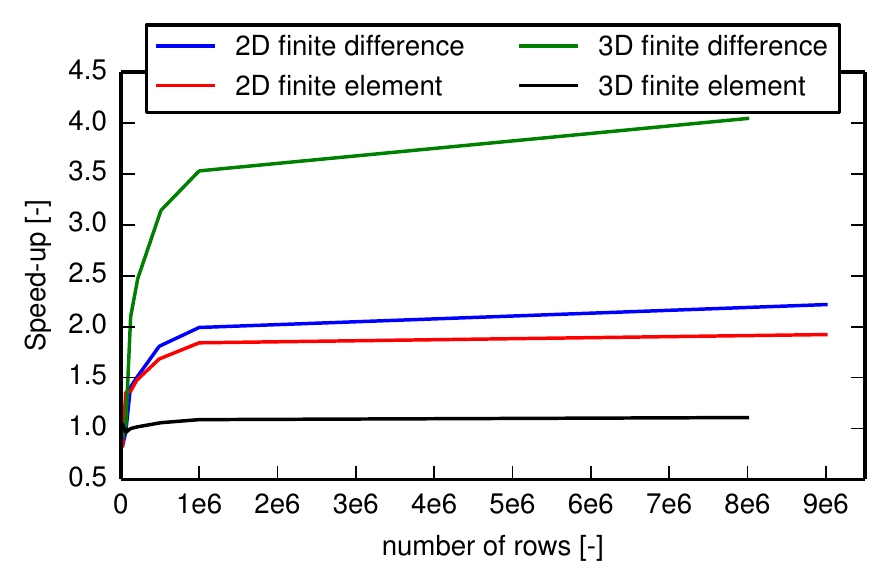}
    \caption{Speed-up achieved using structured SpMV algorithm over native Kokkos Kernels SpMV algorithm for 2D finite difference (5pts), 2D finite element (9pts), 3D finite difference (7pts) and 3D finite element (27pts) stencils on a NVIDIA V100 GPU}
    \label{fig:structured_spmv}
\end{figure}

In Figure~\ref{fig:structured_spmv} we present the computational speed-up between the structured SpMV approach and the native SpMV implementation in Kokkos Kernels. The comparison is performed using a NVIDIA V100 GPU. The structured SpMV provides good performance compared to the native implementation, in the case of the 27pts stencil, further improvement could be achieved by reducing occupancy in order to improve reuse and thus improve the achieved speed-up. We compare to the native implementation here as it is slightly faster that vendor kernels for these use cases.

%% file: sparse-spgemm.tex
The target kernel here is $C=AB$, where $A$, $B$ and $C$ are sparse matrices of size $m\times n$, $n \times k$, and $m \times k$, respectively. 
SpGEMM finds application in graph analysis and scientific computing domains, especially in multigrid solvers such as MueLu~\cite{muelu}.

\subsubsection{Standard SpGEMM Kernel}
\label{sec:usualspgemm}

Our implementation~\cite{MehmetSpGEMM2017,MehmetSpGEMM2018} follows Gustavson's algorithm~\cite{Gustavson}, which iterates over the rows of $A$.
For each row $A(i,:)$, the rows of $B$ which correspond to the column indices of the nonzeros in $A(i,:)$ are accumulated to obtain the resulting row $C(i,:)$ as

\begin{equation}
C(i, :) = \sum_{j \in A(i,:)} A (i,j)B(j,:).
\label{eq:spgemm}
\end{equation}

Note that the number of nonzeros in $C(i,:)$ is not known in advance.

Our implementation consists of two major phases: symbolic and numeric. 
The symbolic phase computes the size of $C$, that is, the number of nonzeros in each row of $C$, whereas the numeric phase computes the column indices and the values of those nonzeros.
In the overall SpGEMM implementation, we first allocate the \textit{row pointers} of $C$ (also known as \textit{row offsets} in the CRS format).
We may compress $B$ into $B_C$ by representing the column indices by bits instead of integers.
Then we perform the symbolic phase on $A$ and $B_C$ to fill the row pointers array.
The last entry of the row pointers array amounts to the total number of nonzeros in $C$, i.e., $nnz(C)$.
Then we allocate two more arrays to store the \textit{column indices} and \textit{values} of $C$, each of which is size of $nnz(C)$.
Finally, we perform the numeric phase on $A$ and $B$ to fill the indices and values arrays.

The symbolic and numeric phases both use an \emph{accumulator} to keep track of the entries in each row of $C$. The accumulator may be sparse or dense, depending on the parallel back-end and input matrices.

In the sparse case, we use a data structure called HashmapAccumulator that stores key-value pairs $(k, v)$. On insertion of $(k_{new}, v_{new})$, $k_{new}$ is inserted if not already present, and the value is accumulated: $v := \mathit{accum}(v, v_{new})$. For the symbolic phase with compression, $\mathit{accum}$ is a bitwise OR. For the numeric phase, it is addition. SpGEMM uses a two-level HashmapAccumulator, where level-1 ($L_1$) and level-2 ($L_2$) components reside in the fast (shared) and slow (global) memory, respectively. We first attempt to insert the entries into $L_1$. If it is full, we lazily create $L_2$ and insert the remaining entries into $L_2$.

Our dense accumulator is a simple array - a bit vector for symbolic, and a scalar array for numeric. Because of the high memory cost of a dense scalar array per thread, the dense numeric accumulator is only used on CPUs.

We have different parallelization methods which vary in terms of how they exploit the hierarchical parallelism provided by Kokkos.
Depending on the input matrix properties, the best method is picked and used for the entire computation.
On the matrices with fewer FLOPs per row, we assign a row block to each team where each row within the block is assigned to a thread.
The rows of $B$ referenced by the current row of $A$ are processed sequentially and the entries in each referenced row of $B$ are processed using vector parallelism.
On the matrices with more FLOPs, we assign individual rows to the teams where each thread processes roughly equal amount of nonzeros in the flattened two nested loops using vector parallelism.
The reader is referred to \cite{MehmetSpGEMM2017} and \cite{MehmetSpGEMM2018} for a more detailed description of the implementation and the performance results.

\subsubsection{Jacobi-Fused SpGEMM Kernel}
\label{sec:jacobispgemm}

This kernel implements a Jacobi operation in the form of $C=(I-\omega D^{-1}A)B$, where $A$, $B$ and $C$ are sparse matrices of sizes $m\times m$, $m \times k$, and $m \times k$, respectively. 
Note that $A$ is square and we assume that all of its diagonal entries have nonzero values.
$\omega$ is a scalar, $I$ is the identity matrix, and $D^{-1}$ represents the inverse of the diagonal matrix of $A$.
This operation is frequently used during the prolongator smoothing in algebraic multigrid solvers~\cite{muelu}.

$C=(I-\omega D^{-1}A)B$ could be implemented by calling the following three kernels: 
\begin{enumerate}
    \item a usual SpGEMM in the form of $E=AB$,
    \item scaling in the form of $F=D^{-1}E$, and
    \item matrix addition in the form of $C=B-\omega F$,
\end{enumerate}
where $E$ and $F$ are intermediate matrices. 
However, in our Jacobi-fused implementation, we fuse Jacobi operation into the computations associated with the $i$th row of $A$ (equivalently, $E$, $F$, and finally $C)$ in the numeric phase of SpGEMM described in Section~\ref{sec:usualspgemm}.
Note that the sparsity patterns of $E$ and $C$ are the same because of the nonzero diagonal entries assumption about $A$.
Therefore, Jacobi-fused SpGEMM uses the usual symbolic kernel to compute the number of nonzeros in $C$.

Consider the $i$th row $E(i,:)$ of the matrix in the first step, which corresponds to $\sum_{A(i,j)\neq 0}A(i,j)B(j,:)$ according to (\ref{eq:spgemm}).
Here, we fuse the Jacobi operation as 
\begin{equation}
    C(i,:)=B(i:)-\omega D^{-1}(i)E(i,:),
\end{equation}
where $D^{-1}(i)$ denotes the $i$th entry of $D^{-1}$.
This means in addition to the usual SpGEMM computation that obtains row $E(i,:)$, the Jacobi-fused kernel also
\begin{itemize}
    \item inserts the entries of row $B(i,:)$ to the hashmap accumulator and
    \item multiplies each $E(i,j)$ entry with scalar $-\omega D^{-1}(i)$.
\end{itemize}
The FLOP count of the fused kernel is smaller than the non-fused implementation by $m-1$ multiply operations, because the fused kernel multiplies $\omega$ and $D^{-1}(i)$ only once for each row.
More importantly, the fused approach launches only one kernel instead of three kernels, hence reduces the kernel launch latency.
Furthermore, it eliminates the costs associated to creating and maintaining the intermediate matrices $E$ and $F$, which may also involve additional communication steps at the distributed-memory level.

\begin{table}[t!]
\caption{Performance comparison of the non-fused (MSAK: multiply-scale-add kernel) and fused (KK: Kokkos Kernels) Jacobi implementations on an MPI+Cuda build of Trilinos~\cite{trilinos-website}.}
\begin{center}
 \begin{tabular}{lrrrr} 
 \toprule
 & & \multicolumn{2}{c}{runtime (ms)} & \\ \cmidrule(lr){3-4} 
 matrix & \#GPUs & MSAK & KK & KK/MSAK \\ \midrule
 ecology1	&4	&45	&27	&0.59\\
	&16	&37	&24	&0.65\\
	&64	&33	&22	&0.67\\ \midrule
 dielFilter	&4	&438	&398	&0.91\\
	&16	&233	&209	&0.90\\
	&64	&138	&100	&0.73\\ \midrule
 Queen\_4147	&4	&184	&128	&0.70\\
	&16	&128	&100	&0.78\\
	&64	&117	&97	&0.83\\
 \bottomrule
 \end{tabular}
 \end{center}
 \label{t:jacobifusedspgemm}
\end{table}

Table~\ref{t:jacobifusedspgemm} displays the performance comparison of the non-fused (MSAK) and fused (KK) approaches on three different problems using an MPI+Cuda build of Trilinos~\cite{trilinos-website}.
The experiments were performed using the spectral graph partitioning tool Sphynx~\cite{sphynx} with a multigrid (MueLu~\cite{muelu}) preconditioner used in the LOBPCG eigensolver.
In Table~\ref{t:jacobifusedspgemm}, we only report the running time of the Jacobi operation in the finest level of MueLu, which means that $A$ corresponds to the adjacency matrix of the input graph and $B$ is computed by MueLu as a prolongator matrix.
We obtained the test graphs/matrices from the SuiteSparse matrix collection~\cite{Davis2011}.
ecology1, dielFilterV2real, and Queen\_4147 have 1.0M, 1.2M, and 4.1M rows, and 5.0M, 48.5M, and 329.5M nonzeros, respectively.
The experiments were performed on Summit, in which each node is equipped with six NVIDIA Volta V100 GPUs and two IBM POWER9 CPUs all connected together with NVLink and the interconnect topology is a non-blocking fat tree.
In our Trilinos build, we used CUDA 10 (version 10.1.243), GCC (version 7.4.0), Netlib’s LAPACK (version 3.8.0), and Spectrum MPI (version 10.3.1.2) and set the environment variable CUDA\_LAUNCH\_BLOCKING. In addition, we disabled cuSPARSE and Cuda-aware MPI as suggested by Sphynx~\cite{sphynx}.
Since Trilinos requires a one-to-one mapping of MPI processes and GPUs, the second column of the table corresponds to the number of MPI processes as well.

As seen in Table~\ref{t:jacobifusedspgemm}, the fused KK kernel is 9\%-41\% faster than the non-fused MSAK implementation.
Note that the performance improvement increases as the number of GPUs increases on dielFilterV2real.
This can be explained by the increasing importance of the kernel launch latency and MPI-related overheads as the computation loads of MPI processes get smaller.   
On the contrary, the improvement decreases on ecology1 and Queen\_4147, which could be attributed to the reduction in the FLOP count obtained by the fused kernel getting smaller as the sizes of the (sub)matrices handled by the MPI processes get smaller.

%% file: sparse-spadd.tex
Sparse matrix addition computes $C = \alpha A + \beta B$, where A, B and C are in CRS format. A trivial way to compute this sum is to simply concatenate each row of A and B. However, it is desirable to merge all entries in C with the same row and column so that its size in memory is minimized. Like SpGEMM, SpAdd has two phases: a ``symbolic'' phase that prepares to compute $C$ based on the sparsity patterns of $A$ and $B$, and a fast ``numeric'' phase that can compute $C$ for any $A,B$ as long as their sparsity patterns are the same as when symbolic was called. Both phases have two versions: one for sorted input, where the entries in each row of $A$ and $B$ are ordered by column, and one for unsorted input. Both versions can handle unmerged inputs; if row $i$ of $A$ contains many entries for column $j$, these entries will be additively merged into a single entry $C(i,j)$.

\subsubsection{SpAdd: Unsorted Inputs}
\label{sec:spadd}
\begin{algorithm}
\caption{SpAdd: Symbolic for Unsorted Inputs}
\label{alg:UnsortedSymbolicSpAdd}
\begin{algorithmic}[1]
\small
    \Procedure{UnsortedSymbolic}{Arow, Brow}
        \ForAll{$i, col \in Arow$}
            \State $Atuples \gets (col, A, i)$
        \EndFor
        \ForAll{$i, col \in Brow$}
            \State $Btuples \gets (col, B, i)$
        \EndFor
        \State $Ctuples \gets $ sort $Atuples \cup Btuples$ by column
        \State $count \gets -1$
        \ForAll{$(col, mat, i) \in Ctuples$} \LineComment{Count unique entries}
            \If {$col$ was not seen before}
                $count \gets count + 1$
            \EndIf
            \If {$mat = A$}
                \State $Apos[i] \gets count$
            \Else
                \State $Bpos[i] \gets count$
            \EndIf
        \EndFor
    \LineComment{Include the final entry} 
    \State\Return{$count+1$}
    \EndProcedure
\end{algorithmic}
\end{algorithm}

Algorithm \ref{alg:UnsortedSymbolicSpAdd} describes the unsorted symbolic phase within each row. It has two goals: count the entries in $C$, and compute scatter indices for each entry in $A$ and $B$. The most performance-critical component is sorting. On GPUs, a team-parallel bitonic sort is used (Section \ref{sec:utils}), while on CPUs, a serial radix sort is used. Like SpGEMM, a parallel prefix sum computes $C$'s row offsets from counts. The numeric phase is simple and fast - the final sum is produced by scattering $A$'s entries using $Apos$, and $B$'s entries using $Bpos$, directly to the correct index within the row of $C$.

\begin{figure}[t]
    \centering
    \includegraphics[scale=0.9]{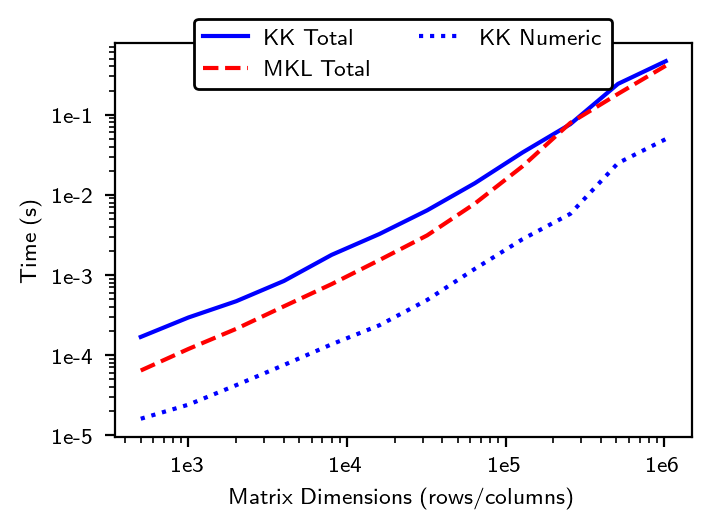}
     \vspace{-0.3cm}
    \caption{SpAdd on unsorted inputs: Kokkos Kernels vs. MKL}
    \label{fig:spadd_unsorted}
     \vspace{-0.5cm}
\end{figure}

Figure \ref{fig:spadd_unsorted} compares against MKL for unsorted inputs. Here, MKL is about 2x faster up to about 100,000 rows, and Kokkos Kernels nearly catches up for larger inputs. However, the Kokkos Kernels numeric phase is on average 6.8x faster than MKL's overall SpAdd.

\subsubsection{SpAdd: Sorted Inputs}
For sorted inputs, computing each row of $C$ is equivalent to merging two sorted lists. On CPUs, this merging is done by the trivial sequential algorithm. The symbolic phase only computes the length of the merged list. For the numeric phase, $C$'s entries and values have been allocated to the correct size. The merged list is computed again, this time storing both the column indices and summed values to the correct offsets in these arrays.

On GPUs, it is profitable to exploit SIMD parallelism within each row. The sorted list merging is done by forming a bitonic sequence in shared memory: $A$'s entries, followed by $B$'s entries in reverse. The merging phase of bitonic sort can sort this sequence in $\lceil log_2(n_A + n_B) \rceil$ parallel steps. As with the unsorted SpAdd algorithm, the sorting also permutes the source index of each entry in $A$ or $B$ in order to form $Apos$ and $Bpos$. The numeric phase is then identical to the unsorted inputs case.

\begin{figure}[t]
    \centering
    \includegraphics[scale=0.9]{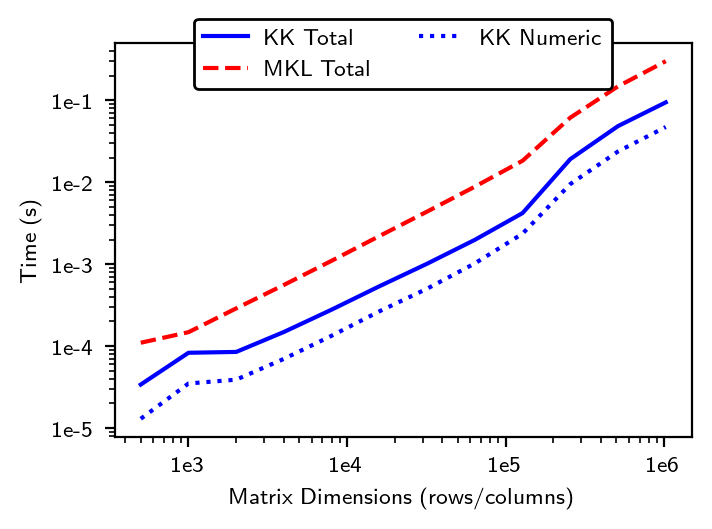}
     \vspace{-0.3cm}
    \caption{SpAdd on sorted inputs: Kokkos Kernels vs. MKL}
    \label{fig:spadd_sorted}
     \vspace{-0.3cm}
\end{figure}
Figure \ref{fig:spadd_sorted} compares Kokkos Kernels with the Intel MKL for SpAdd on sorted inputs. The input matrices are square and each have 30 randomized entries per row. 8 OpenMP threads on a Broadwell Xeon CPU were used. MKL's interface assumes all inputs are unsorted, and it does not  separate the symbolic and numeric phases. On average, Kokkos Kernels gives a 3.6x speedup overall, and a 7.4x speedup for numeric only.

\begin{figure}[t]
    \centering
    \includegraphics[scale=0.9]{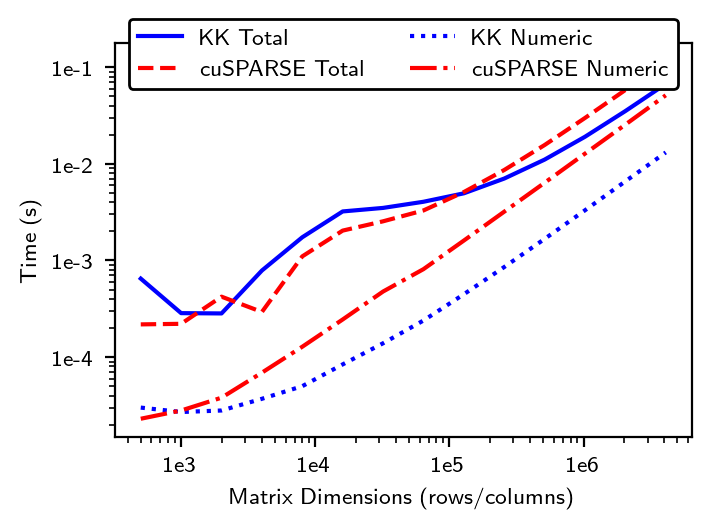}
     \vspace{-0.3cm}
    \caption{SpAdd on sorted inputs: Kokkos Kernels vs. cuSPARSE}
    \label{fig:spadd_sorted_cuda}
     \vspace{-0.5cm}
\end{figure}

Figure \ref{fig:spadd_sorted_cuda} compares Kokkos Kernels with cuSPARSE on an NVIDIA V100. cuSPARSE requires the inputs to be sorted, but it does separate the symbolic and numeric phases. For inputs with fewer than 10,000 rows, cuSPARSE is faster overall, but for inputs larger than 200,000 rows Kokkos Kernels is between 23\% and 70\% faster. Distributed sparse solvers typically use more than 10,000 rows per node because strong scaling breaks down, so in practice Kokkos Kernels should be faster \cite{PETSc_Scaling}. The numeric phase (symbolic reuse) is up to 4x faster for Kokkos Kernels.

%% file: sparse-sptrsv.tex
Kokkos Kernels provides sparse triangular solve (SpTRSV) capabilities to solve linear systems $L*x=b$ or $U*x=b$ for $x$, where $L$ is a lower, and $U$ is an upper, non-singular triangular matrix and $x$ and $b$ are vectors. Sparse triangular solve kernels are an important component for numerical linear algebra algorithms, for example to complete the direct solution to an LU factorized linear system, or with preconditioning using incomplete LU factorization for iterative methods like multigrid or conjugate gradient.

Available SpTRSV options include a default implementation and API, with optional vendor support for NVIDIA’s cuSPARSE implementation for use on NVIDIA GPUs, and a collection of algorithms that take advantage of a matrix’s supernodal structure (referred to as the supernode-based SpTRSV). All algorithms are written using Kokkos and Kokkos Kernels, allowing portability across many-core architectures. Each algorithm is split into separate symbolic and solve phases, where the symbolic phase is responsible for computing and storing information used to expose maximal parallelism to the solve phase. The symbolic phase results can be reused while the matrix structure remains unchanged, improving performance and efficiency of the solve phase for cases with repeat solves as in preconditioned iterative methods.

\subsubsection{Sparse Triangular Solver for incomplete LU}

The default SpTRSV algorithm options employ methods described in Maxim Naumov’s paper ~\cite{naumov2011parallel}. In the basic level scheduling algorithm, the symbolic phase analyzes the matrix structure on the host, performing level scheduling to group matrix rows that can be computed independently into level sets within a dependency tree. The solve phase launches a solve kernel for each level set and employs Kokkos’ hierarchical parallelism, parallelizing solve computation across each row within a level set to take advantage of the parallelism exposed by the symbolic phase.

For cases where the solve kernel’s time is dominated by kernel launch overheads, due to the levels with too few rows to merit any parallelism, a variant of the basic level scheduling algorithm called “chaining” can be used. This variant can be useful particularly for GPU runs, where consecutive levels of the dependency tree that have few rows and little work (or with number of rows falling under a set threshold) are “chained” together into a single kernel launch. On the GPU, this requires that the "chained" kernel launch run on a single thread-block to utilize GPU hardware to synchronize between levels; though this is inefficient usage of the GPU, the reduction in accumulated kernel launch overhead times may improve solve performance, for example with banded matrices where all rows have dependencies. The advantage of the portable implementation of Naumov \emph{et al}. \cite{naumov2011parallel} is to provide a portable preconditioning option on other GPUs.

\subsubsection{Sparse Triangular Solver for direct LU}
\label{sec:sn_sptrsv}

We have also developed a sparse triangular solver that takes advantage of
the special sparsity structures of the triangular matrices that come from
the direct factorization of a sparse matrix~\cite{Yamazaki:2020}.
Specifically, these triangular matrices often has
a set of consecutive rows or columns with similar sparsity patterns,
which can be grouped into a \emph{supernodal} block.
Instead of processing the columns of the sparse triangular matrix,
our supernode-based triangular solver operates on these supernodal blocks.
Though we may be explicitly storing zeros to form the supernodal blocks
and operating with these zeros, this supernode-based solver can
reduce the number of kernels launches and expose the hierarchical parallelism
on top of a traditional parallel triangular solve algorithm (e.g. level scheduling).

Our implementation of the supernode-based triangular solver
uses several key features of Kokkos Kernels including the team-level BLAS kernels.
In particular, at each level of scheduling,
we launch a single functor, where a team of threads execute
a sequence of team-level BLAS kernels on each of the independent supernodal blocks, of variable supernode sizes,
in parallel
(e.g., \textsc{trsm} to compute the solution block with the diagonal block,
followed by \textsc{gemv} to update the remaining blocks with the corresponding off-diagonal blocks).
Hence, the team-level kernels can effectively map the supernodes to
the hierarchical parallelism available on the GPU.

The dense triangular-solve \textsc{trsm}, which is needed to compute a solution block,
is a fundamentally sequential algorithm, and the team of threads may not be able to exploit the parallelism
as well as for the matrix-vector multiply \textsc{gemv} used to update the solution vector.
To avoid the potential performance bottleneck with \textsc{trsm},
we provide an option to explicitly invert the diagonal blocks
(using batched \textsc{trtri})
in the numeric phase. 
We then use the team-level \textsc{gemv}, instead of \textsc{trsm},
to compute each of the solution blocks in the solve phase.
Moreover, we can apply the inverse of the diagonal
blocks to the corresponding off-diagonal blocks in the numeric
phase, using our batched \textsc{trmm} routine 
(new nonzero entries are introduced only within the blocks).
Then, in the solve phase, we combine two \textsc{gemv} calls
(one to compute the solution block and the other to update the remaining solutions)
into a single \textsc{gemv} call.
This is equivalent to the partitioned inverse~\cite{Alvarado:1993},
which partitions the triangular matrix and expresses its inverse as
the product of the inverse of the partitioned matrices.
We revisited this technique where the triangular matrix is partitioned based on the supernodal level set,
and demonstrated that it is a valid option on a GPU.
Table~\ref{tab:sptrsv} compares our SpTRSV performance with that of cuSPARSE on an NVIDIA P100 GPU when SuperLU and METIS ordering are used to compute the LU factors.

\begin{table}[t]
\label{tab:sptrsv}
\caption{Kokkos Kernels and cuSPARSE SpTRSV solve time in seconds (lower + upper triangular solve), using SuperLU and METIS (NVIDIA P100)}
\begin{tabular}{l r r r r}
\toprule
matrix       & $n$        & $nnz/n$ & cuSPARSE       & Kokkos Kernels \\ \hline
tmt\_sym     & 726,713    & 7.0     & 0.021 + 0.031  & 0.005 + 0.010  \\
af\_shell7   & 504,855    & 34.8    & 0.034 + 0.047  & 0.005 + 0.012  \\
ecology2     & 999,999    & 4.0     & 0.026 + 0.037  & 0.009 + 0.022  \\
thermal2     & 1,228,045  & 7.0     & 0.022 + 0.030  & 0.008 + 0.015  \\
\bottomrule
\end{tabular}
\end{table}

%% file: softwaredesign.tex
\section{Software Design}\label{sec:software_design}
Kokkos Kernels provides a header-only C++ template library for instantiating linear algebra and graph kernels. 
Kokkos Kernels still requires linking against an installed library (shared or static) due to Kokkos core,  third-party libraries and explicit template instantiation.
Like Kokkos itself, Kokkos Kernels provides single-source, multiple-platform C++ code through compile-time dispatch to different code paths by changing execution policy and memory space input types.
The final executable code is generated when building downstream applications, not Kokkos Kernels itself, which creates a so-called transitive dependency problem. Our goals in the design are:

\begin{itemize}
    \item Limiting expensive template-heavy compile times through explicit instantiation (called \textit{eti-kernels})
    \item Transparently dispatching to vendor-optimized third-party libraries when available without needing to modify user code (called \textit{tpl-kernels})
    \item Enabling use of kernels not covered through \textit{eti-kernels} and \textit{tpl-kernels} (called \textit{inline-kernels})
\end{itemize}
We use Kokkos \texttt{View} for all our interfaces.
One problem with this design is that different types can actually represent the exact same data with the same properties, see frame below, which leads to a combinatorial explosion of different signatures for an otherwise identical function. 
\begin{lstlisting}[language=C++,frame=single]
T norm(View<T*>(...));
T norm(View<const T*>(...));
T norm(View<const T[150]>(...));
T norm(View<T*,LayoutLeft>(...));
T norm(View<T*,LayoutRight,DefaultMemSpace>(...));
T norm(View<const T*,LayoutRight,Unmanaged>(...));
\end{lstlisting}


There are three issues because of this flexibility: 
\begin{itemize}
  \item Instantiating a function simply for \texttt{View$<$double*$>$} will not cover a call for \texttt{View$<$const double*$>$}.
  \item While a function, \texttt{norm(View$<$double*$>$)} for example, can be mapped to a \textit{tpl-kernel} an equally allowed one, \texttt{norm(View$<$double*, LayoutStride$>$)} for instance, might not.
  \item Without further measures calling the otherwise identical functions above, will result in six compiled variants of the \texttt{norm} function, with the associated increase in compile time and binary size.
\end{itemize}

To enable both the transparent dispatch to \textit{tpl-kernels},  \textit{eti-kernels} and \textit{inline-kernels} and resolving these issues the software design is deploying a two layer implementation. 
The user-facing \textit{dispatch-layer} consists of functions templated on the Kokkos \texttt{View} types.
These functions are always implicitly instantiated by the user.
They first unify the types of the \texttt{View} arguments as much as possible and then dispatch to the \textit{implementation-layer} which consists of partial specializations of classes.

In order to then dispatch from the same function to either the \textit{inline-kernel}, \textit{tpl-kernel} or \textit{eti-kernel} helper arguments are introduced for the parameters of the implementation layer.
They provide the information about whether a specific kernel variant was explicitly instantiated or can map to a third party library. 

\begin{lstlisting}[language=C++,frame=single]
template<class ViewType>
auto norm(ViewType a) {
  // get the common type 
  CommonVT<ViewType>::type common_a = a;
  
  // Call the implementation layer
  return NormImpl<CommonVT,
    TPL_Variant_Exists<CommonVT>::value,
    ETI_Variant_Exists<CommonVT>::value>::norm(common_a);
}
\end{lstlisting}

Inside an application the call to \texttt{norm} will now only inline instantiate the actual implementation of the kernel, if both \texttt{TPL\_Variant\_Exists$<$CommonVT$>$} and \texttt{ETI\_Variant\_Exists$<$CommonVT$>$} are false.

Furthermore, even the instantiations for which \texttt{TPL\_Variant\_Exists$<$CommonVT$>$} is true can still distinguish between different \texttt{View} types.
That allows Kokkos Kernels to target a TPL for host data (e.g. Intel's MKL) and for GPU data (e.g. NVIDIA's CuBLAS and CuSparse) in the same executable, since the memory type information is part of the \texttt{View} type. 

During compilation of the Kokkos Kernels library, a list can be specified for which \textit{eti-kernels} shall be generated.
For the actual generation the same generic code is then used which applications would hit when they implicitly instantiate a kernel (i.e. \texttt{ETI\_Variant\_Exists$<$CommonVT$>$} is simply always set to false during library compilation).

Another use case introduced later in the development of Kokkos Kernels, is the ability for users to choose at runtime between different variants.
This can be used both for choosing between different algorithms for the same function implemented by Kokkos Kernels itself and choosing between calling the \textit{tpl-kernel}, \textit{inline-kernel} or \textit{eti-kernel}. 
The user provides that type of information via a \textit{handle} as an argument to the \textit{dispatch-layer} function.
Choosing between different algorithms in the Kokkos Kernel implementation happens in the \textit{implementation-layer}.
However, choosing between \textit{tpl-kernel}, \textit{inline-kernel} and \textit{eti-kernel} happens at the \textit{dispatch-layer} to avoid unnecessary explicit instantiations of the \textit{inline-kernel}s during compilation of the \textit{tpl-kernel}s.

\begin{lstlisting}[language=C++,frame=single]
template<class ViewType>
auto func(KKHandle h, ViewType a) {
  // get the common type 
  CommonVT<ViewType>::type common_a = a;
  
  // Choose between kokkos-kernels and tpl variant
  if(h.no_tpl_kernel) {
    // Call the implementation layer with TPL false
    return FuncImpl<CommonVT,
      false,
      ETI_Variant_Exists<CommonVT>::value>::func(h, common_a);
  } else {
    // Call the implementation layer
    return FuncImpl<CommonVT,
      TPL_Variant_Exists<CommonVT>::value,
      ETI_Variant_Exists<CommonVT>::value>::func(h, common_a);
  }
}
\end{lstlisting}

\section{Conclusion}
We described the capabilities of the Kokkos Kernels library. We also showed how interface design could help us achieve better performance. We showed examples for reusing symbolic information, team-level routines for batched linear algebra and team-level utilities like hashing and sorting will help us achieve portable performance. We demonstrated several new results for sparse/dense matrix kernels and graph kernels. We have also compared the portable kernels with state-of-the-art native kernels when applicable. Kokkos Kernels continues to grow in terms of its capabilities. We plan to fully support upcoming accelerators from Intel and AMD in the near future.

%% file: main.bbl
\begin{thebibliography}{10}
\providecommand{\url}[1]{#1}
\csname url@samestyle\endcsname
\providecommand{\newblock}{\relax}
\providecommand{\bibinfo}[2]{#2}
\providecommand{\BIBentrySTDinterwordspacing}{\spaceskip=0pt\relax}
\providecommand{\BIBentryALTinterwordstretchfactor}{4}
\providecommand{\BIBentryALTinterwordspacing}{\spaceskip=\fontdimen2\font plus
\BIBentryALTinterwordstretchfactor\fontdimen3\font minus
  \fontdimen4\font\relax}
\providecommand{\BIBforeignlanguage}[2]{{%
\expandafter\ifx\csname l@#1\endcsname\relax
\typeout{** WARNING: IEEEtran.bst: No hyphenation pattern has been}%
\typeout{** loaded for the language `#1'. Using the pattern for}%
\typeout{** the default language instead.}%
\else
\language=\csname l@#1\endcsname
\fi
#2}}
\providecommand{\BIBdecl}{\relax}
\BIBdecl

\bibitem{edwards2014kokkos}
H.~C. Edwards, C.~R. Trott, and D.~Sunderland, ``Kokkos: Enabling manycore
  performance portability through polymorphic memory access patterns,''
  \emph{Journal of Parallel and Distributed Computing}, vol.~74, no.~12, pp.
  3202--3216, 2014.

\bibitem{lawson1979basic}
C.~L. Lawson, R.~J. Hanson, D.~R. Kincaid, and F.~T. Krogh, ``Basic linear
  algebra subprograms for {F}ortran usage,'' \emph{ACM Transactions on
  Mathematical Software (TOMS)}, vol.~5, no.~3, pp. 308--323, 1979.

\bibitem{wang2014intel}
E.~Wang, Q.~Zhang, B.~Shen, G.~Zhang, X.~Lu, Q.~Wu, and Y.~Wang, ``Intel {M}ath
  {K}ernel {L}ibrary,'' in \emph{High-Performance Computing on the
  Intel{\textregistered} Xeon Phi™}.\hskip 1em plus 0.5em minus 0.4em\relax
  Springer, 2014, pp. 167--188.

\bibitem{naumov2010cusparse}
M.~Naumov, L.~Chien, P.~Vandermersch, and U.~Kapasi, ``{CUSPARSE} library,'' in
  \emph{GPU Technology Conference}, 2010.

\bibitem{trilinos-website}
T.~{T}rilinos~{P}roject {T}eam.

\bibitem{balay2019petsc}
S.~Balay, S.~Abhyankar, M.~Adams, J.~Brown, P.~Brune, K.~Buschelman, L.~Dalcin,
  A.~Dener, V.~Eijkhout, W.~Gropp \emph{et~al.}, \emph{{PETSc} users manual},
  Argonne National Laboratory, 2019.

\bibitem{falgout2002hypre}
R.~D. Falgout and U.~M. Yang, ``hypre: A library of high performance
  preconditioners,'' in \emph{International Conference on Computational
  Science}.\hskip 1em plus 0.5em minus 0.4em\relax Springer, 2002, pp.
  632--641.

\bibitem{rocsparse}
\BIBentryALTinterwordspacing
AMD. (2021, feb) roc{SPARSE} {D}ocumentation. AMD. [Online]. Available:
  \url{https://rocsparse.readthedocs.io/en/master/}
\BIBentrySTDinterwordspacing

\bibitem{Intel_MKL}
\BIBentryALTinterwordspacing
Intel. (2021, feb) Intel {oneAPI} {M}ath {K}ernel {L}ibrary ({oneMKL}) - {D}ata
  {P}arallel {C++} {D}eveloper {R}eference. Intel. [Online]. Available:
  \url{https://software.intel.com/content/www/us/en/develop/documentation/oneapi-mkl-dpcpp-developer-reference/top.html}
\BIBentrySTDinterwordspacing

\bibitem{Thrust}
\BIBentryALTinterwordspacing
N.~Bell and J.~Hoberock, ``Thrust: A {P}roductivity-{O}riented {L}ibrary for
  {CUDA},'' in \emph{GPU Computing Gems Jade Edition}, ser. Applications of GPU
  Computing Series, W.~mei W.~Hwu, Ed.\hskip 1em plus 0.5em minus 0.4em\relax
  Boston: Morgan Kaufmann, 2012, pp. 359--371. [Online]. Available:
  \url{https://www.sciencedirect.com/science/article/pii/B9780123859631000265}
\BIBentrySTDinterwordspacing

\bibitem{merrill2015cub}
D.~Merrill, ``{CUB},'' \url{https://nvlabs.github.io/cub}, 2015.

\bibitem{spmv_report}
S.~D. Hammond and C.~R. Trott, ``Optimizing the performance of sparse-matrix
  vector products on next-generation processors,'' Sandia National
  Laboratories, Albuquerque, NM, Tech. Rep. SAND2017-5953, 2017.

\bibitem{muelu}
L.~Berger-Vergiat, C.~A. Glusa, J.~J. Hu, C.~Siefert, R.~S. Tuminaro, M.~Mayr,
  A.~Prokopenko, and T.~Wiesner, ``{MueLu User's Guide.}'' 1 2019.

\bibitem{MehmetSpGEMM2017}
M.~{Deveci}, C.~{Trott}, and S.~{Rajamanickam}, ``Performance-portable sparse
  matrix-matrix multiplication for many-core architectures,'' in \emph{2017
  IEEE International Parallel and Distributed Processing Symposium Workshops
  (IPDPSW)}, 2017, pp. 693--702.

\bibitem{MehmetSpGEMM2018}
\BIBentryALTinterwordspacing
M.~Deveci, C.~Trott, and S.~Rajamanickam, ``Multithreaded sparse matrix-matrix
  multiplication for many-core and {GPU} architectures,'' \emph{Parallel
  Computing}, vol.~78, pp. 33--46, 2018. [Online]. Available:
  \url{https://www.sciencedirect.com/science/article/pii/S0167819118301923}
\BIBentrySTDinterwordspacing

\bibitem{Gustavson}
F.~G. Gustavson, ``Two fast algorithms for sparse matrices: Multiplication and
  permuted transposition,'' \emph{ACM Transactions on Mathematical Software
  (TOMS)}, vol.~4, no.~3, pp. 250--269, 1978.

\bibitem{sphynx}
S.~{Acer}, E.~G. {Boman}, and S.~{Rajamanickam}, ``{SPHYNX}: {S}pectral
  {P}artitioning for {HY}brid a{N}d a{X}elerator-enabled systems,'' in
  \emph{2020 IEEE International Parallel and Distributed Processing Symposium
  Workshops (IPDPSW)}, 2020, pp. 440--449.

\bibitem{Davis2011}
\BIBentryALTinterwordspacing
T.~A. Davis and Y.~Hu, ``The {U}niversity of {F}lorida sparse matrix
  collection,'' \emph{ACM Trans. Math. Softw.}, vol.~38, no.~1, pp. 1:1--1:25,
  Dec. 2011. [Online]. Available:
  \url{http://doi.acm.org/10.1145/2049662.2049663}
\BIBentrySTDinterwordspacing

\bibitem{PETSc_Scaling}
M.~Lange, G.~Gorman, M.~Weiland, L.~Mitchell, and J.~Southern, ``Achieving
  {E}fficient {S}trong {S}caling with {PETSc} {U}sing {H}ybrid {MPI/OpenMP}
  {O}ptimisation,'' in \emph{Supercomputing}, J.~M. Kunkel, T.~Ludwig, and
  H.~W. Meuer, Eds.\hskip 1em plus 0.5em minus 0.4em\relax Berlin, Heidelberg:
  Springer Berlin Heidelberg, 2013, pp. 97--108.

\bibitem{naumov2011parallel}
M.~Naumov, ``Parallel solution of sparse triangular linear systems in the
  preconditioned iterative methods on the {GPU},'' \emph{NVIDIA Corp.,
  Westford, MA, USA, Tech. Rep. NVR-2011}, vol.~1, 2011.

\bibitem{Yamazaki:2020}
\BIBentryALTinterwordspacing
I.~Yamazaki, S.~Rajamanickam, and N.~Ellingwood, ``Performance portable
  supernode-based sparse triangular solver for manycore architectures,'' in
  \emph{49th International Conference on Parallel Processing - ICPP}, ser. ICPP
  '20.\hskip 1em plus 0.5em minus 0.4em\relax New York, NY, USA: Association
  for Computing Machinery, 2020. [Online]. Available:
  \url{https://doi.org/10.1145/3404397.3404428}
\BIBentrySTDinterwordspacing

\bibitem{Alvarado:1993}
F.~L. Alvarado, A.~Pothen, and R.~Schreiber, ``Highly parallel sparse
  triangular solution,'' in \emph{Graph Theory and Sparse Matrix Computation.
  The IMA Volumes in Mathematics and its Applications}, A.~G. A, J.~R. Gilbert,
  and J.~W.~H. Liu, Eds.\hskip 1em plus 0.5em minus 0.4em\relax New York, NY:
  Springer, 1993, ch.~56, pp. 141--157.

\bibitem{phipps2017embedded}
E.~Phipps, M.~D'Elia, H.~C. Edwards, M.~Hoemmen, J.~Hu, and S.~Rajamanickam,
  ``Embedded ensemble propagation for improving performance, portability, and
  scalability of uncertainty quantification on emerging computational
  architectures,'' \emph{SIAM Journal on Scientific Computing}, vol.~39, no.~2,
  pp. C162--C193, 2017.

\bibitem{Nvidia_cuBLAS}
\BIBentryALTinterwordspacing
NVIDIA. (2021, feb) {CUDA} {T}oolkit {D}ocumentation. NVIDIA. [Online].
  Available: \url{https://docs.nvidia.com/cuda/cublas/index.html}
\BIBentrySTDinterwordspacing

\bibitem{MAGMA_Batched}
A.~Abdelfattah, A.~Haidar, S.~Tomov, and J.~Dongarra, ``Novel {HPC} techniques
  to batch execution of many variable size {BLAS} computations on {GPU}s,'' in
  \emph{Proceedings of the International Conference on Supercomputing}, ser.
  ICS '17, 06 2017.

\bibitem{kim2017designing}
K.~Kim, T.~B. Costa, M.~Deveci, A.~M. Bradley, S.~D. Hammond, M.~E. Guney,
  S.~Knepper, S.~Story, and S.~Rajamanickam, ``Designing vector-friendly
  compact {BLAS} and {LAPACK} kernels,'' in \emph{Proceedings of the
  International Conference for High Performance Computing, Networking, Storage
  and Analysis}, 2017, pp. 1--12.

\bibitem{MehmetColoring}
M.~{Deveci}, E.~G. {Boman}, K.~D. {Devine}, and S.~{Rajamanickam}, ``Parallel
  graph coloring for manycore architectures,'' in \emph{2016 IEEE International
  Parallel and Distributed Processing Symposium (IPDPS)}, 2016, pp. 892--901.

\bibitem{D2Coloring}
M.~K. {Taş}, K.~{Kaya}, and E.~{Saule}, ``Greed is good: Parallel algorithms
  for bipartite-graph partial coloring on multicore architectures,'' in
  \emph{2017 46th International Conference on Parallel Processing (ICPP)},
  2017, pp. 503--512.

\bibitem{BellMIS2}
N.~Bell, S.~Dalton, and L.~Olson, ``\BIBforeignlanguage{English (US)}{Exposing
  fine-grained parallelism in algebraic multigrid methods},''
  \emph{\BIBforeignlanguage{English (US)}{SIAM Journal of Scientific
  Computing}}, vol.~34, no.~4, pp. C123--C152, 2012.

\bibitem{cusp}
\BIBentryALTinterwordspacing
S.~Dalton, N.~Bell, L.~Olson, and M.~Garland, ``{CUSP}: Generic {P}arallel
  {A}lgorithms for {S}parse {M}atrix and {G}raph {C}omputations,'' 2014,
  version 0.5.0. [Online]. Available: \url{http://cusplibrary.github.io/}
\BIBentrySTDinterwordspacing

\end{thebibliography}
